\documentclass[pdflatex,sn-mathphys]{sn-jnl}% Math and Physical Sciences Reference Style
\jyear{2023}%

\theoremstyle{thmstyleone}%
\theoremstyle{thmstyletwo}%

\theoremstyle{thmstylethree}%

\usepackage{caption}
\usepackage{subcaption}
\begin{document}

\title[Article Title]{Bottom-Up Approach to Explore Alpha-Amylase Assisted Membrane Remodelling}


\author[1]{\fnm{Harshit} \sur{Kumar}}
\equalcont{These authors contributed equally to this work.}
\author[1]{\fnm{Sayar} \sur{Mandal}}
\equalcont{These authors contributed equally to this work.}
\author[1]{\fnm{Suhasi} \sur{Gupta}}
\author[1]{\fnm{Hemraj} \sur{Meena}}
\author[1]{\fnm{Mayur} \sur{Kadu}}
\author[1]{\fnm{Rajni} \sur{Kudawla}}
\author[2]{\fnm{Pratibha} \sur{Sharma}}
\author[2]{\fnm{Indu Pal} \sur{Kaur}}
\author*[3]{\fnm{John H} \sur{Ipsen}}\email{ipsen@memphys.sdu.dk}
\author*[1]{\fnm{Tripta} \sur{Bhatia}}\email{triptabhatia@iisermohali.ac.in}

\affil[1]{\orgdiv{Department of Physical Sciences}, \orgname{Indian Institute of Science Education and Research Mohali}, \orgaddress{\street{SAS Nagar}, \city{Knowledge City}, \postcode{140306}, \state{Punjab}, \country{India}}}

\affil[2]{\orgdiv{University Institute of Pharmaceutical Sciences}, \orgname{Punjab University}, \orgaddress{\street{Sector 14,
Chandigarh}, \city{City}, \postcode{160014}, \state{Punjab}, \country{India}}}

\affil[3]{\orgdiv{Department of Physics, Chemistry and Pharmacy}, \orgname{University of Southern Denmark}, \orgaddress{\street{Campusvej}, \city{Odense}, \postcode{5230 M}, \country{Denmark}}}




\abstract{
Soluble alpha-amylases play an important role in the catabolism of polysaccharides. In this work, we show that the enzyme can interact with the lipid membrane and further alter its mechanical properties. Vesicle fluctuation spectroscopy is used for quantitative measurement of the membrane bending rigidity of phosphatidylcholines lipid vesicles from the shape fluctuation based on the whole contour of Giant Unilamellar Vesicles (GUVs). The bending rigidity of the lipid vesicles of 1-palmitoyl-2-oleoyl-sn-glycero-3-phosphocholine in water increases significantly with the presence of 0.14 micromolar alpha-amylase in the exterior solution. However, as the concentration increases above 1 micromolar, the bending rigidity decreases but remains higher than estimated without the protein. Contact between the alpha-amylase in the outer solution and the outer leaflet leads to spontaneous membrane curvature and the corresponding morphological changes of the GUVs. The presence of outbuds directly demonstrates that AA has a preferable interaction with the membrane, giving a positive spontaneous curvature of $C_0 \leq 0.05  \ \mu \rm{m}^{-1}$ at $18 \ \mu$M $(\approx$ 1 mg/ml) of AA concentration. Above 1 mg/ml of AA concentration the shape of GUVs collapse completely suggesting a highly convoluted state.}

\maketitle

\section*{Introduction}\label{sec1}
\begin{figure}[ht!]%
\centering
\includegraphics[width=.8\linewidth]{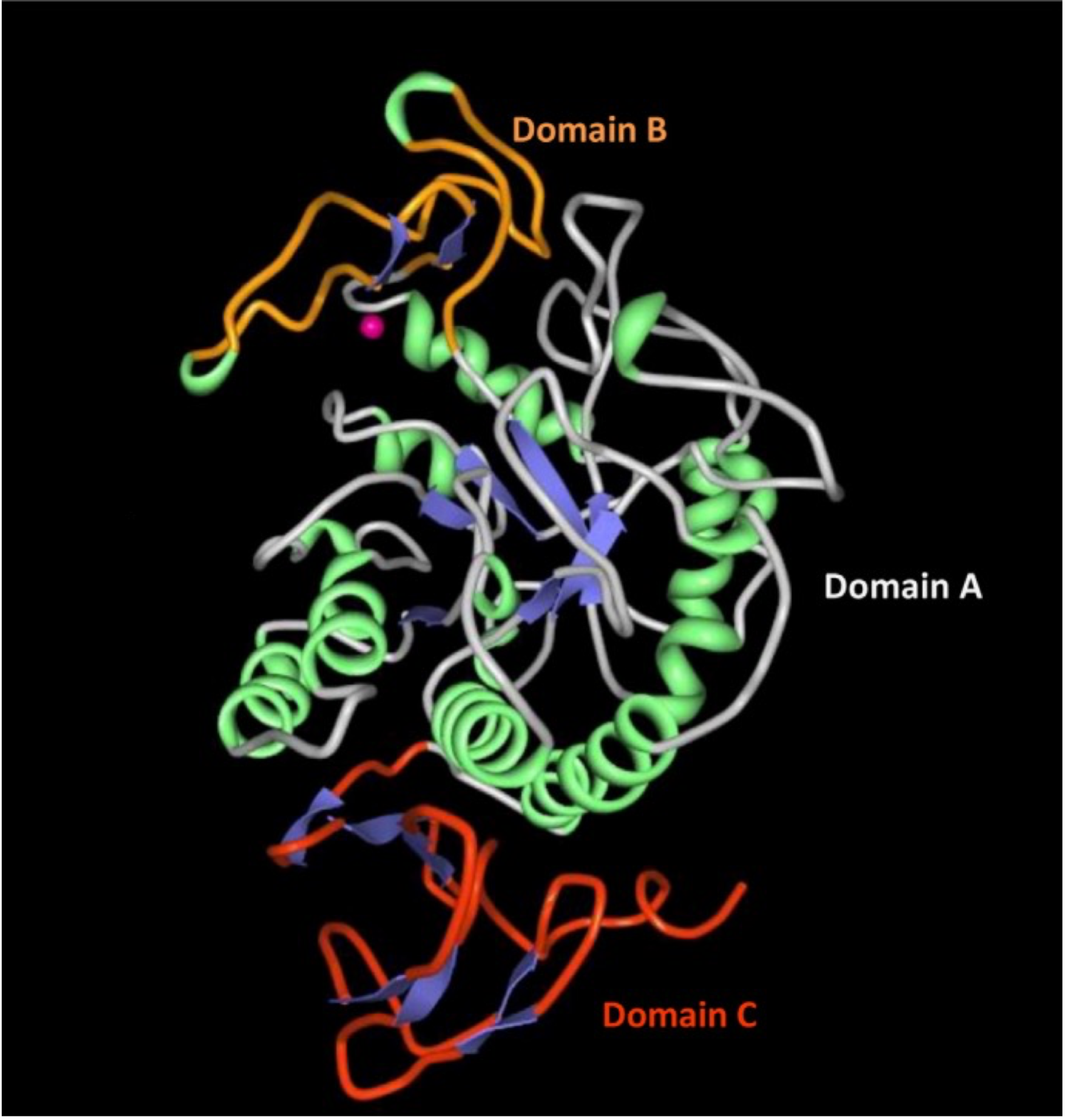}
\caption{The $\alpha$-amylase structural model, image was downloaded from the Protein data bank. The pink dot indicates the Ca$^{2+}$-binding site. The blue and green parts represent the beta strands and alpha helixes, respectively. Domains A–C are set apart by different colors and labeled by the matched color.\cite{a7}}
\label{amylases}
\end{figure}

Alpha amylases ($\alpha$-amylases, AA) play an important role in catabolism in biology by facilitating the breakdown of large $\alpha$-bonded polysaccharides such as starch into short-chain carbohydrates available for metabolism. AA is documented as a crucial component in bacterial biofilm degradation and thus plays a crucial role in wound healing. \cite{b77,b76} Antibiofilm property of $\alpha$ -amylase is reported against pathogens such as {\em Vibrio cholerae} and drug resistant pathogens such as methicillin resistant {\em Staphylococcus aureus} (MRSA) and {\em Pseudomonas aeruginosa}. The enzyme degrades the biofilm by disrupting the exopolymeric saccharide (EPS) matrix by acting on its glycosidic linkage. \cite{b78,b79} AA are found both in soluble forms and in membrane-bound homologues. For example, saliva $\alpha$ -amylase (sAA) is soluble without known membrane affinity, while pancreatic $\alpha$ - amylase (pAA) is membrane bound with a lipid anchor, or {\em Bacillus subtilis} synthesizes soluble and membrane bound forms of $\alpha$ -amylases \cite{b70}. Figure \ref{amylases} shows the structural model of AA. The colloidal stability of AA in solutions is primarily facilitated by net charges (positive under neutral pH conditions) \cite{b72}, often providing a highly crowded environment for enzyme activity, while aggregation is found at high pH-values ($>8$)  and high concentrations. Soluble AA naturally binds to polysaccharide fibers \cite{b74, b75}, while the literature does not indicate interactions with lipid membranes.\\
Wolfgang Helfrich proposed a simple model for the description of curvature-elastic deformation of the shape of the lipid membrane 50 years ago \cite{b10}. Helfrich's model proved to be very powerful in describing a plethora of phenomena involving membrane conformation, and it laid the foundation for many theoretical predictions of phenomena involving membrane conformations and experimental applications. This seminal paper emphasized the prominent role of conformational fluctuations in membranes. We prepare cell models, Giant Unilammellar Vesicles (GUVs) that are analyzed by vesicle fluctuation spectroscopy for the bending rigidity measurements. Next we discuss parameter and error estimates by combining experiments and theory. The resulting membrane bending rigidity of POPC membranes in the presence of AA is discussed, which also includes comparative measurements with common pertubants of membrane mechanical properties. The theory and experiments to obtain estimates of membrane asymmetry are given, together with the discussion and conclusions.
\parshape=0

\paragraph {Theory of Vesicle fluctuations}
It was the subsequent work of Brochard and Lennon \cite{b0} that showed the potential to analyze conformational fluctuations of the micrograph membrane to obtain an estimate of the model parameters of the Helfrich model. Vesicle conformations are well described by the Helfrich free energy, which is a general formulation that takes the form \cite{b10}.
\begin{equation}
F_{\text{Hel}} = \oint_{S} \left[ \frac{\kappa}{2} \left(2 H - C_0 \right)^2 \right] dA + \sigma  A 
\label{Helfrich}
\end{equation}
\noindent
for a single vesicle with spherical topology. The properties of the material are characterized by three parameters: the elastic modulus of bend $(\kappa)$, the tension of the internal membrane $(\sigma)$, and the spontaneous curvature $(C_0)$.\cite{b31} $H, \ A$ represent the mean curvature and the total surface area of the GUV, respectively. In general, it is very difficult to describe conformational fluctuations based on Eq.(\ref{Helfrich}), but for small fluctuations around simple reference shapes, it is possible, as described below.
\paragraph {Quasi-spherical membranes}
For many experimental vesicles, a realistic approximation of conformational fluctuations is to consider small deviations from the spherical shape.
\begin{equation}
\label{xv}
    \mathbf{X}(\theta,\phi,t) = R [1 + u(\theta, \phi,t)]\mathbf{n} (\theta, \phi)
\end{equation}
\noindent
where $(\theta, \phi)$ are the spherical angles and $u(\theta, \phi,t)$ represent the relative deviations of a radius of the sphere $R$ and $\mathbf{n}$ is the normal vector of the unit sphere. \cite{b18} The quasi-spherical approximate form can naturally be decomposed into spherical harmonics $(Y_l^n)$ where $l,n$ are the mode numbers.
\begin{equation}
u(\theta, \phi) =  \sum_l \sum_n u_{ln} Y_l^n(\theta,\phi) 
\end{equation}
which corresponds to the Fourier decomposition on a flat background. \cite{b43,b44} Eq.(1) can now be expressed by Eq.(3) and take a relatively simple form when the terms up quadratic order in $u_{ln}$ are kept.
Employing the equipartition theorem gives the following \cite{b18} 

\begin{equation}
	\langle u_{ln} u^{*}_{l'n'} \rangle  = \frac{ k_B T / \kappa}   { (l+2)(l-1) [ l (l+1) + \Sigma ] }  \  \delta_{ll'} \delta {n n'}
\label{sphequipartition}
\end{equation}

%\langle u_{ln} u^{*}_{l'n'} \rangle  = \left \big [ \frac{ k_B T} { \kappa}  \frac {1} { (l+2)(l-1) ( l(l+1) + \Sigma ) } \big ] \right \delta_{ll'} \delta {n n'}
 
\begin{equation}
\label{tension}
\Sigma = \frac{\sigma R^2}{\kappa} + \frac{R^2 (C_0)^2}{2}  
	+ 2 C_0 R  
\end{equation}
\noindent 
Eq.(\ref{sphequipartition}) gives the possibility of obtaining estimates of the model parameters from fluctuation data. This analysis is based on linearizing the complex nonlinear analysis Eq.(\ref{Helfrich}), which relates the correlation functions directly to the Hamiltonian. It has been clear that during recent years the above analysis is too simplified and that the estimates of model parameters represent highly renormalized quantities. In particular, the tension parameter $(\Sigma)$ in Eq.(\ref{tension}) obtained from fluctuations has been shown to be the frame tension rather than the internal tension. \cite{b47,b33} 
\section*{Experimental finding of the quasi-spherical fluctuation spectrum}\label{sec3}
\subsection*{Contour detection} 
\begin{figure}[H]
    \centering
    \includegraphics[width=.8\linewidth]{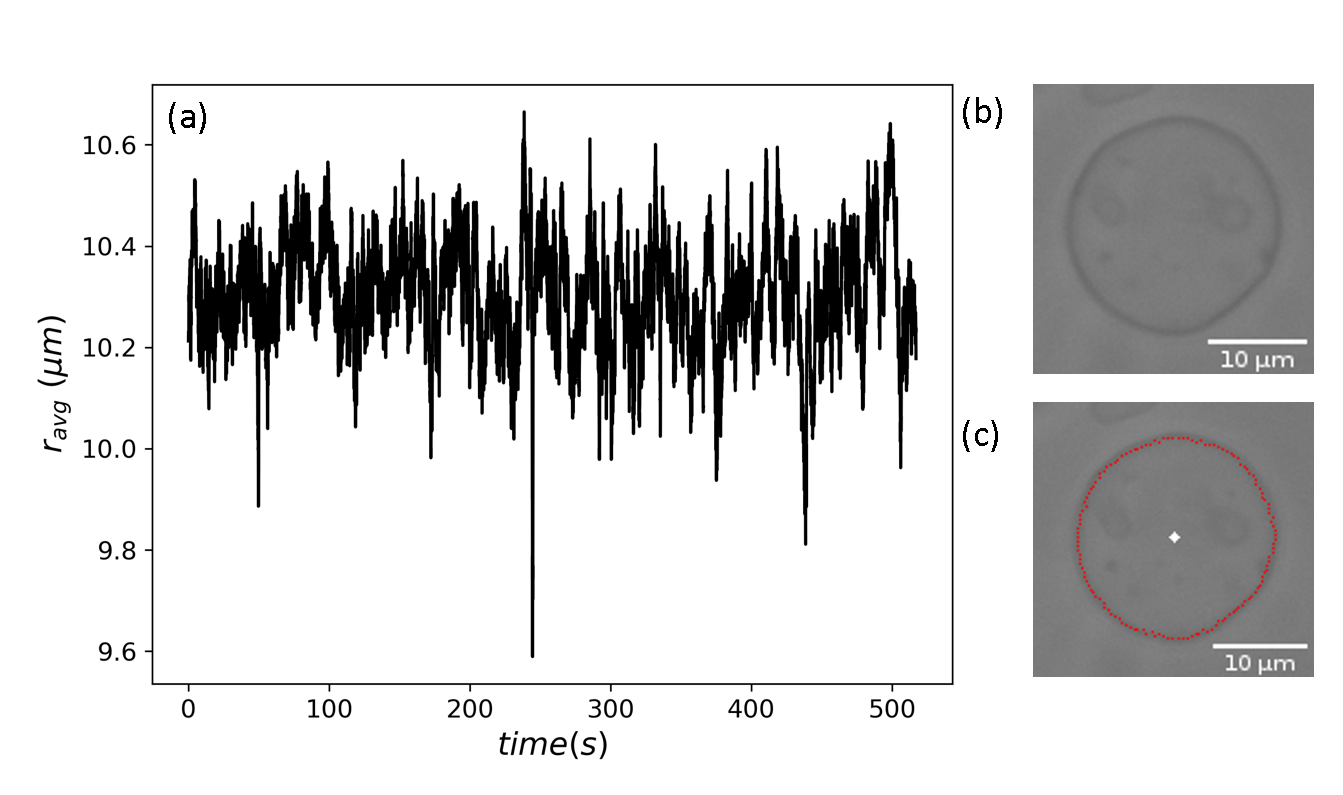}\hspace{1.5cm}
    \caption{GUV contour detection. (a) The contour radius $(r_{\text{avg}})$ of  is plotted with time. (b,c) The vesicle contour and center detected at a given instant of time for the same GUV. POPC GUVs are prepared in 10 mM Hepes.}
    \label{fig:mean_radius}
\end{figure}

\begin{figure}[H]
    \centering
    \includegraphics[width=.8\linewidth]{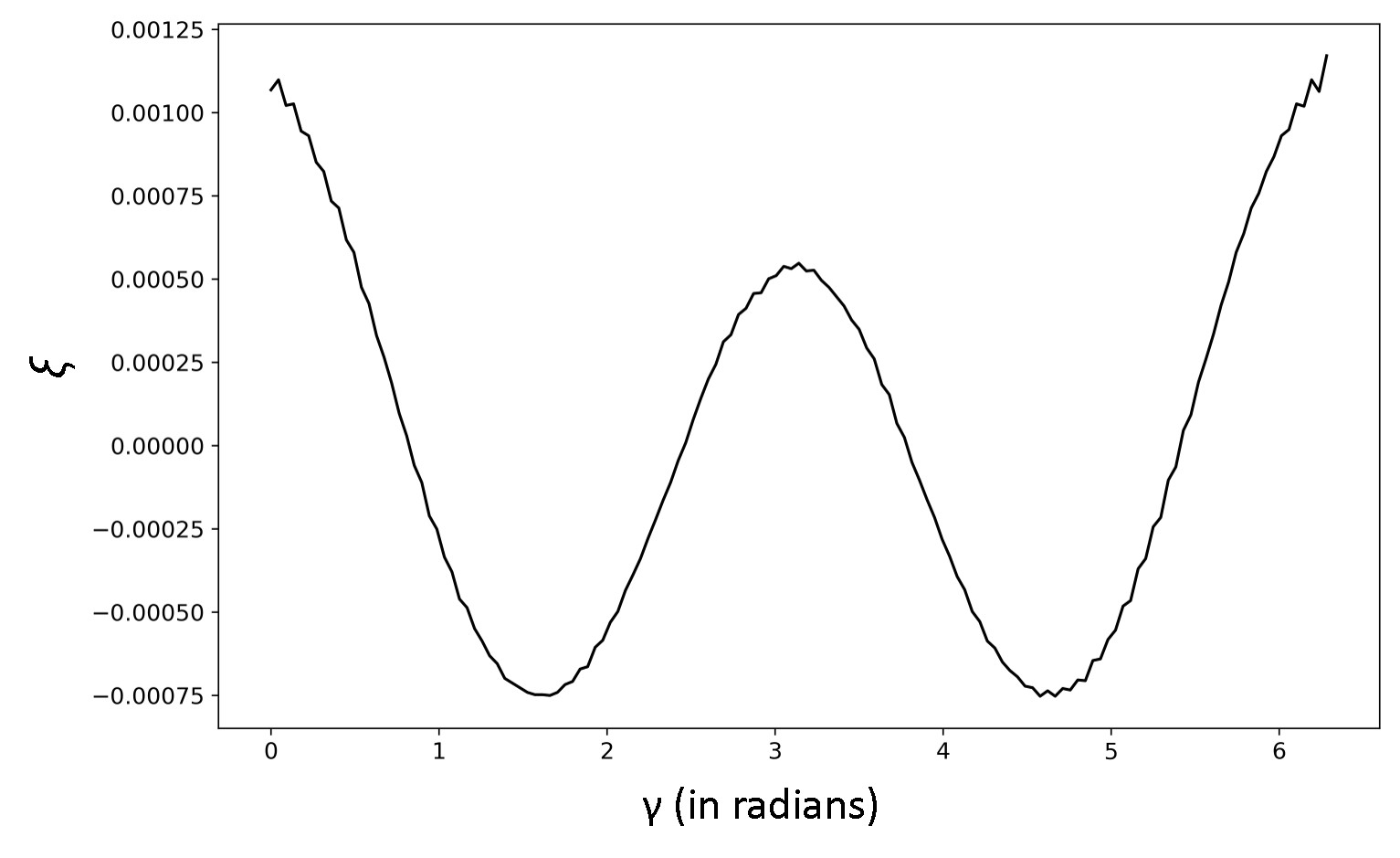}\hspace{1.5cm}
    \caption{Angular Auto-Correlation $\xi(\gamma,t)$ for a single frame is plotted. POPC GUVs are prepared in 10 mM Hepes.}
    \label{fig:angautocorr}
\end{figure}

\begin{figure}[H]
    \centering
    \includegraphics[width=.8\linewidth]{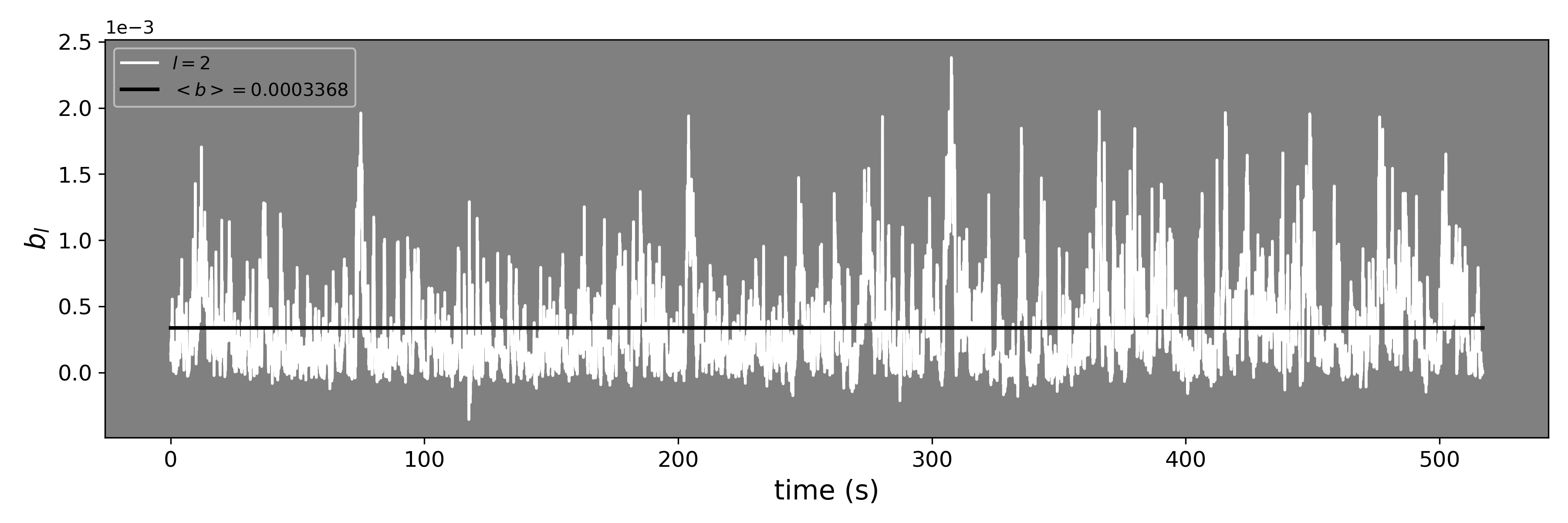}
    \caption{Legendre spectrum for $l=2$ mode is shown. GUV is prepared of POPC lipid in 10 mM Hepes.}
    \label{fig:leg_spectrum}
\end{figure}

For a closed membrane compartment, such as a quasi-spherical giant unilamellar vesicle (GUV), the camera captures the position of the equatorial contour of the GUV membrane. We define for a given frame (acquired at time t at latitude theta) the radius $r_{eq}(\phi,t) = R(1+u(\pi/2,\phi,t))$ and the average radius for the frame. Figure \ref{fig:mean_radius}a shows the time series of the measurement of the average radius of a GUV. Figures \ref{fig:mean_radius}b,c show the instantaneous vesicle contour and the center of the same GUV respectively.
\begin{equation}
   r_{\text{avg}}(t)  =\frac{1}{2\pi}\int_{0}^{2 \pi}  r_{eq}(\phi,t)  d\phi
\end{equation}
\noindent
At the equator $\theta=\pi/2$, a polar coordinate system was used to scan the image for contour. The algorithm requires users to select an approximate center, an approximate average radius, and the width of the region of interest in the first frame of the data. The width of the region of interest is symmetric with respect to the average radius selected by the user. This minimizes information loss near the contour during the transformation of the image from Cartesian to polar coordinates. A radial intensity profile $r_{\text{eq}}  (\phi,t)$ is extracted for the contour points of $\vec{X}_{\text{eq}} (\phi,t)$ from each frame at regular intervals of polar angle ($\phi$). The pixel location of the minima is used as the contour point for dark boundary GUVs. \\
\noindent
The continuous angular auto-correlation function at the equator is given as
\begin{equation}
\label{xi}
    \xi_{eq}(\gamma, t)=\frac{1}{2\pi r_{\text{avg}}(t)^2}\int_{0}^{2\pi}[r_{eq}( \phi+\gamma,t)r_{eq}( \gamma,t)- r_{\text{avg}}(t)^2]d\phi
\end{equation}
\noindent
Ideally, a smooth and symmetric double well should be obtained as the angular auto-correlation function, as shown in Figure \ref{fig:angautocorr} for a given frame as a function of $\gamma$. 
\noindent
The equatorial contour of quasi-spherical vesicles is approximately circular, inviting one to explore the periodicity to decompose the configurations into periodic modes. The angular correlation function is most appropriate for the analysis of fluctuations. The experimental $\xi^{exp}(\gamma, t)$ can easily be obtained from the contours obtained and can be
expressed in terms of $u(\pi/2,\phi,t) \ \colon$
\begin{equation}
	\xi^{exp}(\gamma, t)= \frac{1}{2 \pi } \int _0^{2 \pi} d \phi  \left[ u (\frac{\pi}{2}, \phi + \gamma, t) u (\frac{\pi}{2}, \phi , t) - \bar{u}(t)^2\right] 
	\label{xi0}
\end{equation}
\noindent
where $\bar{u}(t)= \frac{1}{2 \pi}\int_0^{2 \pi} d \phi \   u (\frac{\pi}{2}, \phi , t)$. The corresponding theoretical values of $\xi^{th}(\gamma, t)$ take the form of 

\begin{equation}
\xi^{th}(\gamma, t)=\sum_{l_{1} \geq 0} \sum_{l_{2} \geq 0} \sum_{n\neq 0} u_{l_{1}n} (t) u^{*}_{l_{2}n} (t) Y_{l_{1}n} (\pi/2, \gamma) Y^{*}_{l_{2}n} (\pi/2,0)
\end{equation}
\noindent
The average $\langle \xi^{th}(\gamma) \rangle$ can be obtained by replacing the temporal average with a thermal average (ergodic hypothesis) and employing Eq.(\ref{sphequipartition}). A sum rule for spherical harmonic functions has been applied in Eq.(\ref{e24}). 
\begin{eqnarray}
\label{e24}
	\langle \xi^{th}(\gamma) \rangle & =&\sum_{l_{1} \geq 0}  \sum_{n\neq 0} \langle u_{l_{1}n} u^{*}_{l_{1}n}\rangle \sum_n Y_{l_{1} n} (\pi/2, \gamma) Y^{*}_{l_{1}n} (\pi/2,0)  \nonumber \\
	 & =& \frac{1}{2 \pi} \sum_{ l \geq 0}  \frac{2l +1}{2}\langle \mid u_{l_{1}n} \mid ^2 \rangle P_l(\cos \gamma)  \nonumber \\
& =&  \sum_{ l \geq 0}  b_{l}  P_l(\cos \gamma)
\end{eqnarray}
where $l$ is the mode number. Figure \ref{fig:leg_spectrum} shows the Legendre spectrum for mode number two $(l=2)$ for a POPC GUV. A time average of $b_l$ is plotted in Figure \ref{fig:bexp} with the error bar representing the standard error on the mean (SEM).
This form in Eq.(\ref{e24}) shows that the natural basis for the mode decomposition of the equatorial contours is the Legendre functions $P_l(\cos \gamma)$ in the sense that the amplitudes of $\langle \xi^{th}(\gamma) \rangle$ can be expressed directly by the model parameters. Then we can identify $\langle \xi^{th}(\gamma) \rangle$ and $\langle \xi^{exp} (\gamma) \rangle$ in a fitting procedure. 
\section*{Parameter and error estimations}\label{sec4}
\begin{figure}[H]
    \centering
    \includegraphics[width=.8\linewidth]{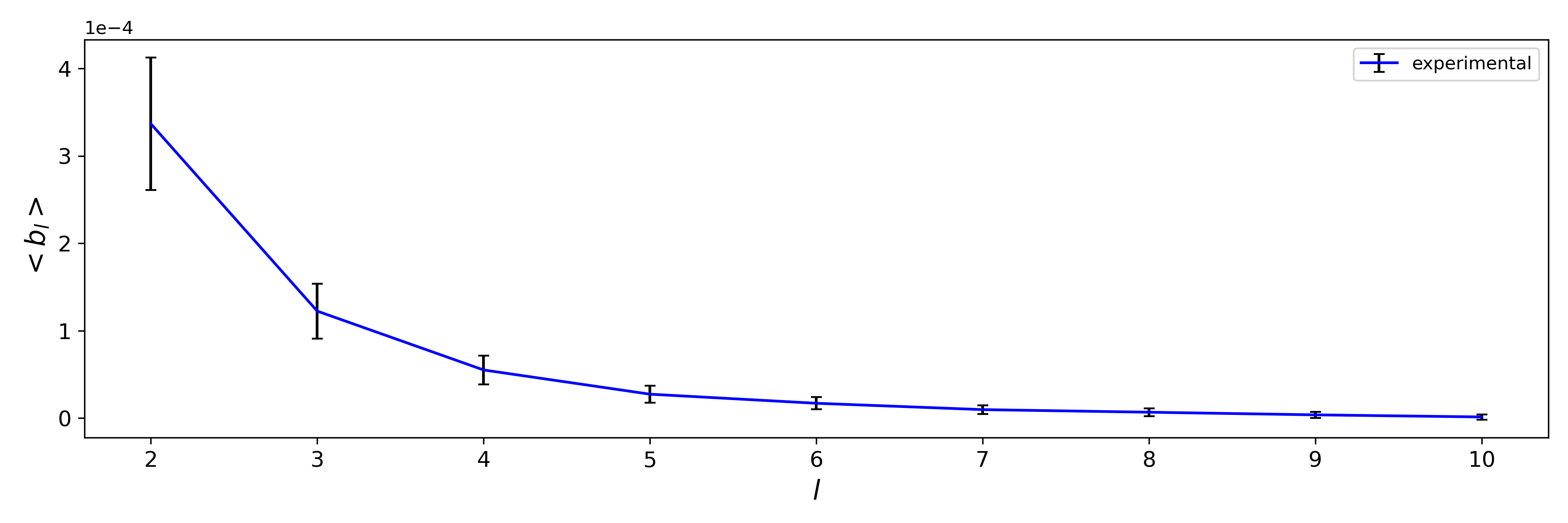}\hspace{1.5cm}
    \caption{Time average Legendre Coefficients for different modes for a single GUV composed of POPC, prepared in 10 mM HEPES.}
    \label{fig:bexp}
\end{figure}

\begin{figure}[H]
    \centering
    \includegraphics[width=.8\linewidth]{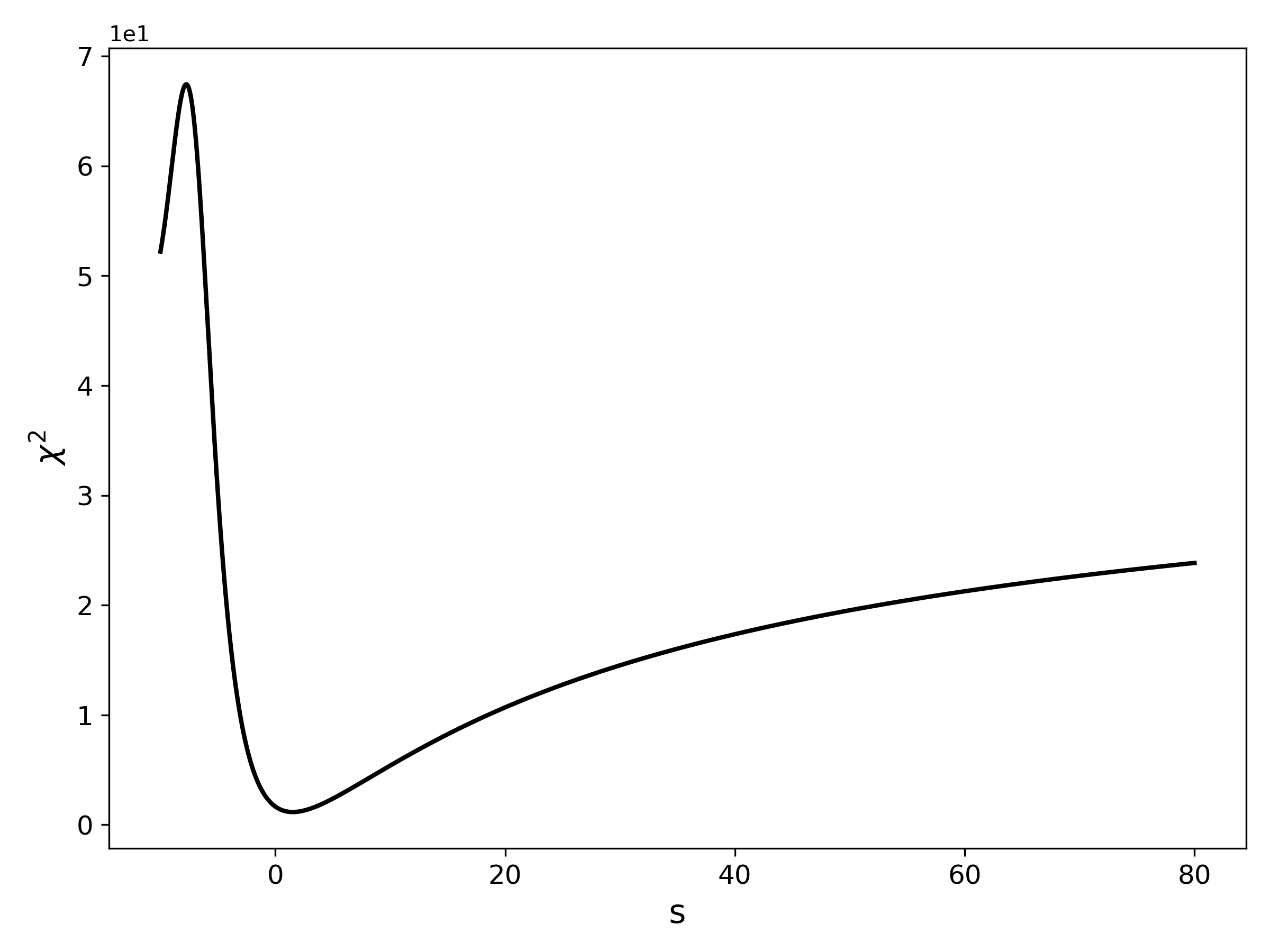}\hspace{1.5cm}
    \caption{Numerical Reconstruction of $\chi^2$ using Eq. \ref{eqn:chi_sq_3} to show the minima in $\chi^2$ for a given value of $s= \Sigma$, the effective surface tension.}
    \label{fig:chi_sq}
\end{figure}

\begin{figure}[H]
    \centering
    \includegraphics[width=.8\linewidth]{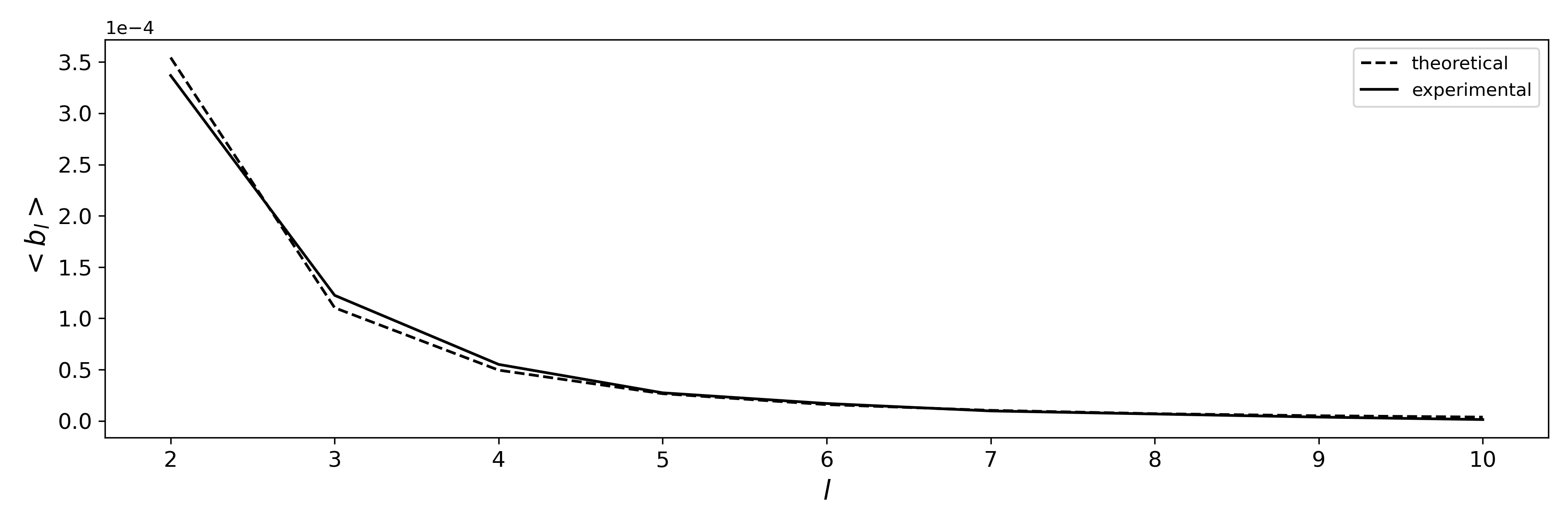}
    \hspace{1.5cm}
    \caption{Fit of the theoretical and experimental mean of Legendre coefficients which is used to estimate $\kappa$.}
    \label{fig:verif}
\end{figure}

Let the theoretical mean $\langle b_l\rangle_{th}\equiv \bar B_l$ and the experimental mean $\langle b_l\rangle_{exp}\equiv B_l$. $\chi^2$ is defined as
\begin{equation}\label{eqn:chi_sq_1}
    \chi^2 = \sum_{l_{min}}^{l_{max}}\left[\frac{B_l-\bar B_l(j,s)}{\sigma_l}\right]^2
\end{equation}
\noindent
where $\sigma_l$ is the standard error in mean, which we obtained for each mode as shown in Figure \ref{fig:bexp}. 
Using Eqs. (\ref{sphequipartition}) and (\ref{e24}), the theoretical mean of the Legendre coefficients is given as
\begin{equation}\label{eqn:B_th}
    \bar B_l(j,s) = \frac{j}{p_l+sq_l}
\end{equation}
where
\begin{align}
    q_l &= \frac{4\pi(l-1)(l+2)}{2l+1} \nonumber \\
    p_l &= \frac{4\pi(l-1)(l+2)(l+1)l}{2l+1}  \nonumber \\
    j &= \left(\frac{k_BT}{\kappa}\right)  \nonumber\\
    s &= \Sigma  \nonumber
\end{align}
\noindent
Substituting $\bar B_l(j,s)$ from Eq. \ref{eqn:B_th} into Eq. \ref{eqn:chi_sq_1} we get,
\begin{equation}\label{eqn:chi_sq_2}
    \chi^2(j,s) = \sum_{l_{min}}^{l_{max}}\left[\frac{p_lB_l-\frac{j}{1+s\frac{q_l}{p_l}}}{p_l\sigma_l}\right]^2
\end{equation}      
\noindent
At the minima of $\chi^2$,
\begin{equation}
    \frac{\partial\chi^2}{\partial j} = -2\sum_{l}\frac{1}{(p_l\sigma_l)^2\left(1+s\frac{q_l}{p_l}\right)}\left(p_lB_l-\frac{j}{1+s\frac{q_l}{p_l}}\right) = 0
\end{equation}
\begin{equation}
    \frac{\partial\chi^2}{\partial s} = 2\sum_{l}\frac{q_l}{p_l}\frac{j}{(p_l\sigma_l)^2\left(1+s\frac{q_l}{p_l}\right)^2}\left(p_lB_l-\frac{j}{1+s\frac{q_l}{p_l}}\right) = 0
\end{equation}
\noindent
The two equations form a set of implicit expressions for the estimators of $\kappa$ and $\Sigma$. Solving for $j$ we get
\begin{equation}
    j(s) = \frac{\sum_l{p_lB_l}/{(p_l\sigma_l)^2\left(1+s\frac{q_l}{p_l}\right)}}{\sum_l{1}/{(p_l\sigma_l)^2\left(1+s\frac{q_l}{p_l}\right)^2}}
    \label{js}
\end{equation}
\noindent
Substituting $j(s)$ in Eq. \ref{eqn:chi_sq_2}, to obtain $\chi^2$ only depending on $s$ and the minimum of $\chi^2$ can be obtained as shown in Figure \ref{fig:chi_sq}. 
\begin{equation}\label{eqn:chi_sq_3}
    \chi^2(j(s),s) = \sum_{l_{min}}^{l_{max}}\left[\frac{p_lB_l-\frac{j(s)}{1+s\frac{q_l}{p_l}}}{p_l\sigma_l}\right]^2
\end{equation}
\noindent
The value of $s$ at the minima of $\chi^2$ is the effective surface tension. Its value is then used to calculate $j(s)$ defined in Eq. \ref{js} and hence the bending rigidity ($\kappa$). Therefore, the obtained $\kappa$ and $s$ are used to obtain the theoretical mean values of the Legendre coefficients. Figure \ref{fig:verif} shows the fit of the theoretical and experimental mean of the Legendre coefficients that give an estimate of the bending rigidity values ($\kappa$) for GUVs. 
\noindent
The covariance matrix $H'$ from Gaussian approximation is given as,
\begin{equation}
    H' =  \begin{bmatrix}
\frac{\partial^2\chi^2}{\partial j^2} & \frac{\partial^2\chi^2}{\partial j\partial s} \\
\\
\frac{\partial^2\chi^2}{\partial j\partial s} & \frac{\partial^2\chi^2}{\partial s^2} 
\end{bmatrix}^{-1}_{min}
\label{hm}
\end{equation}
\noindent
where the components are
\begin{align}
    \frac{\partial^2\chi^2}{\partial j^2} &= 2\sum_l\frac{1}{(p_l\sigma_l)^2\left(1+s\frac{q_l}{p_l}\right)^2}\\
    \frac{\partial^2\chi^2}{\partial j\partial s} &= 2\sum_l\frac{q_l}{p_l}\frac{1}{(p_l\sigma_l)^2\left(1+s\frac{q_l}{p_l}\right)^2}\left(p_lB_l-2\frac{j}{1+s\frac{q_l}{p_l}}\right)\\
    \frac{\partial^2\chi^2}{\partial s^2} &= 2\sum_l \left(\frac{q_l}{p_l}\right)^2\frac{1}{(p_l\sigma_l)^2\left(1+s\frac{q_l}{p_l}\right)^3}\left(\frac{3j}{1+s\frac{q_l}{p_l}}-2p_lB_l\right)
\end{align}
The error in ($\kappa$) is given by $H'_{11}$ and the error in ($s$) is given by $H'_{22}$. \\
$\sigma^2(j) = H'_{11}$ and $\sigma^2(s) = H'_{22}$. Since $j = \frac{k_BT}{\kappa}$ we have $\frac{\sigma^2(j)}{j^2}=\frac{\sigma^2(\kappa)}{\kappa^2}$. Therefore, $\sigma^2(\kappa)=H'_{11}\frac{\kappa^2}{j^2}$.
\section*{Experimental estimation of $\kappa$}\label{sec5}
\begin{figure}[h!]%
\centering
\includegraphics[width=.9\linewidth]{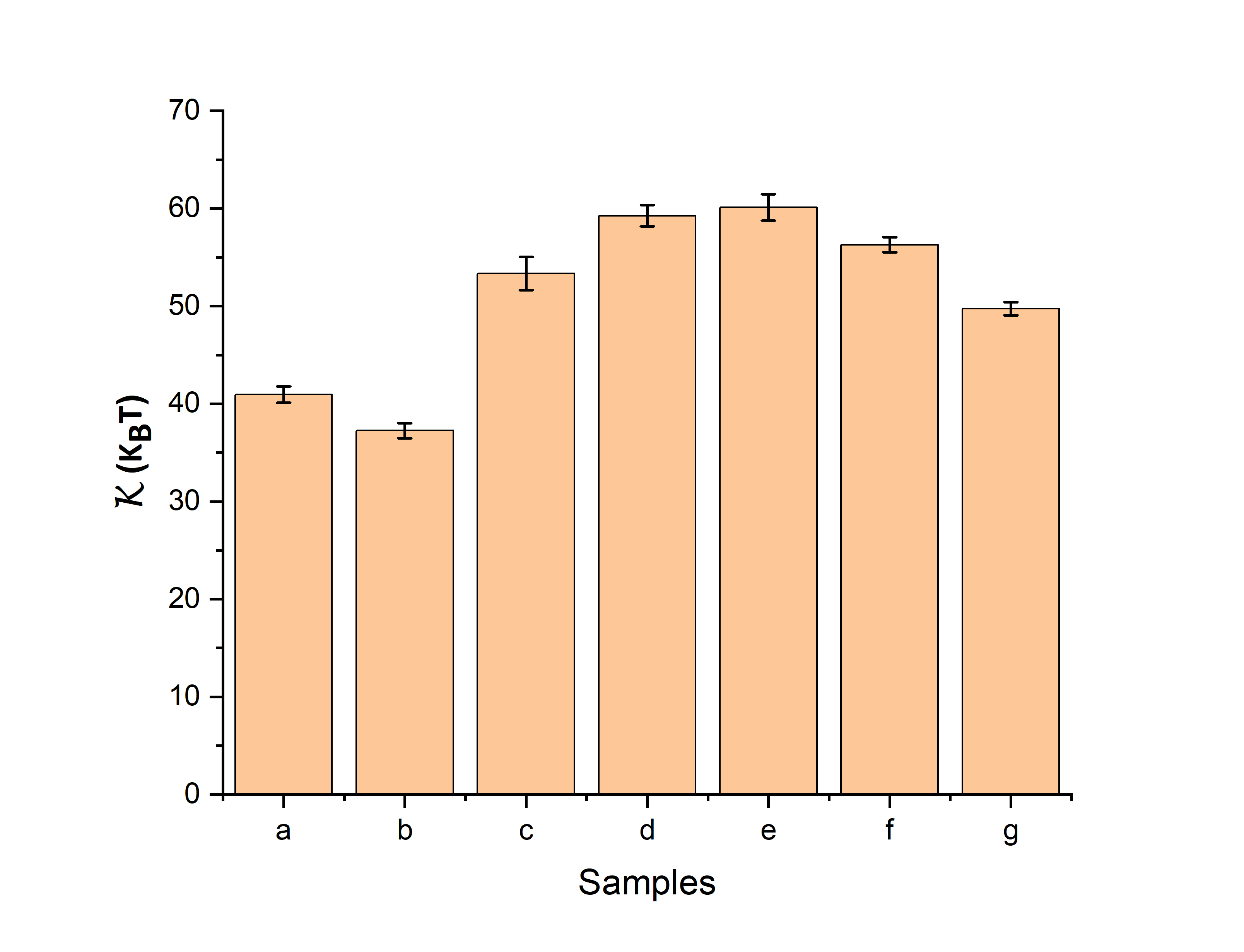 }
\caption{Comparison of bending rigidity $(\kappa)$ values of all POPC GUVs. On the X-axis, label a refers to POPC GUVs in milli-Q water, label b refers to POPC GUVs in 10mM HEPES, label c refers to POPC + cholesterol( 10\%) GUVs in 10mM sucrose, labels d, e, f, g refer to POPC GUVs in milli-Q water with AA concentration (0.14$\mu$M, 1.02$\mu$M, 1.19$\mu$M and 1.34$\mu$M ) respectively. For each sample, we examined at least 10 vesicles from repeated experiments. 
  }\label{popc-samples}
\end{figure}

The GUVs samples are composed of lipids of 1-palmitoyl-2-oleoyl-sn-glycero-3-phosphocholine (POPC) and POPC-cholesterol (10\%) and were prepared in different solutions as listed in Table 1. Figure \ref{popc-samples} shows the estimate of the bending rigidity values ($\kappa$) for GUVs in the absence and presence of AA at different concentrations. We started with a dilute concentration of AA, for example, $0.14 \ \mu$M, and increased it tenfold to $1.34 \ \mu$M for which the value of $\kappa$ is estimated and compared. Furthermore, we compare the effects of AA on the membrane bending rigidity with some well-known perturbances of bending rigidity, such as cholesterol, sugars, and buffer components. \cite{b23,b39,b35,b40,b45,b42,b41,b37,b36,b34}

\begin{table}[t!]
\centering
\begin{tabular}{lrr}
sample-lipids (solution inside and outside) & AA ($\mu$ M) & $\kappa ( k_B T)$ \\
\midrule
a-POPC (water) & 0 & $40.94 \pm 0.83$ \\  
b-POPC (10 mM Hepes) &  0 & $37.25 \pm 0.77$ \\  
c-POPC-chol(10 \%) (10 mM sucrose) & 0 & $53.34 \pm 1.72$ \\
d-POPC (water) & 0.14 & $59.25 \pm 1.08$ \\    
e-POPC (water) & 1.02  & $60.11 \pm 1.36$ \\  
f-POPC (water) & 1.19 & $56.28 \pm 0.77$ \\ 
g-POPC (water) & 1.34 & $49.74 \pm 0.68$ \\
\bottomrule
\end{tabular}
\end{table}
\noindent
To emphasize this important difference between the dilute regime and the 10-fold increase in the concentration regime, we have plotted the estimated value of $\kappa$ for POPC-cholesterol (10\%) GUV prepared in 10 mM sucrose (sample c) in the same figure.  The observed bending rigidities are, in general, in agreement with previous studies of the POPC vesicle \cite{b39}. The presence of HEPES buffer reduces the rigidity of the POPC membrane, cholesterol increases the rigidity of bending \cite{b9,b35}, while small amounts of sugar in solution have a minor effect \cite{b23,b45}. 
The error bars in $\kappa$ shown in Figure \ref{popc-samples} are the standard error on the mean. The error in $\kappa$ of a fluctuating GUV membrane is given by the element of the covariance matrix $H'_{11}$ (as described in Eq. \ref{hm}). We plot all the $\kappa$ values of observed shape fluctuations with their corresponding $H'_{11}$ in the supplementary information.  
\section*{Membrane asymmetry in the presence of the AA}\label{sec6}
\begin{figure}[h!]
        \centering
        \includegraphics[width=.8\linewidth]{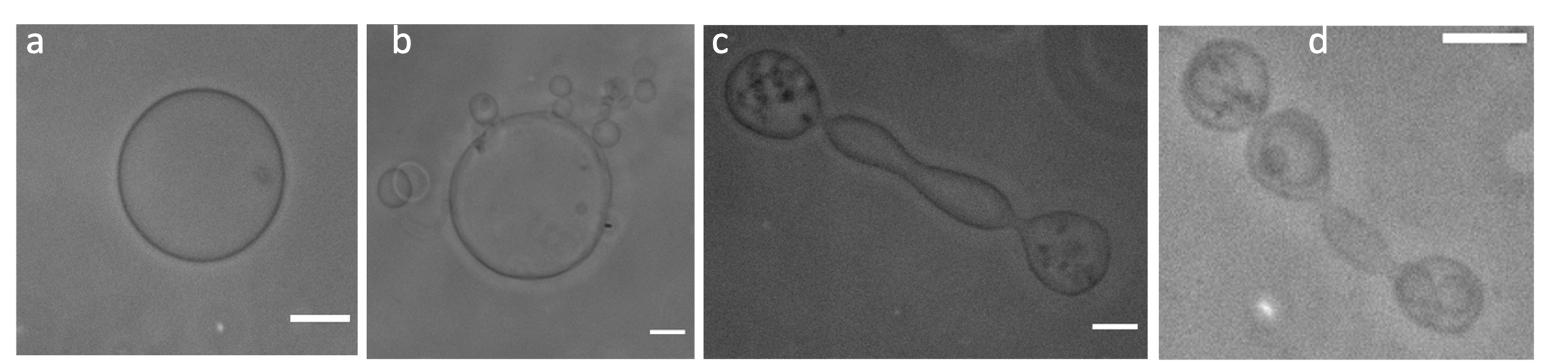}
        \caption{Shape changes of GUVs upon addition of AA in the exterior. (a) A quasi-spherical GUV prepared in water. (b) AA ($\approx 5 \ \mu$M) is added in the exterior compartment, and the image is obtained. Shape changes as observed by phase-contrast microscopy, at (c) $\approx 9.4 \ \mu$M and (c) $\approx 18 \ \mu$M AA concentration are indicated. The buds remain connected to the mother GUV by a narrow membrane neck. The scale bars are $10 \ \mu m$.}
        \label{varyc}
\end{figure}

\begin{figure}[h!]
        \centering
        \includegraphics[width=.8\linewidth]{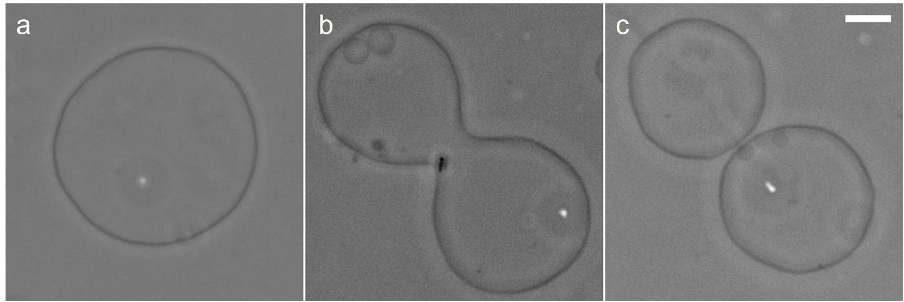}
        \caption{Remodeling of membrane morphology. The first frame shows a quasi-spherical GUV prepared in water. (a) AA ($\approx 18 \ \mu$M) is added in the exterior compartment, and the image is obtained at about $t=104 \ s$. Shape changes, as observed by phase contrast microscopy, evolving from the beginning toward (b) $t=4304 \ s$ and (c) $t=5463 \ s$ are indicated for the same GUV. The buds remain connected to the mother GUV by a narrow membrane neck that gives $C_0 \leq 0.05  \ \mu \rm{m}^{-1}$. The scale bar is $10 \ \mu m$ and applies to all frames. }
        \label{budding}
\end{figure}

Lipid bilayer asymmetry is an extremely important topic in biology. The spontaneous curvature $(C_0)$ defined in Eq. \ref{tension} quantifies membrane asymmetry. \cite{b3,b10,b57,b59,b60,b69} Here, lipid asymmetry refers to a transbilayer membrane asymmetry generated by the asymmetric solution; the outer leaflet is in contact with AA molecules dissolved in water and the inner leaflet is in contact with water without any solute molecules. Because these two leaflets are in contact with different types of solution, the bilayers become asymmetric. \cite{b57,b59,b60,b69} The lipid bilayers of our GUVs contain only a single lipid POPC, which implies that the two leaflets have the same lipid composition and no asymmetry in terms of this composition. \cite{b50,b51} 
\paragraph{In the absence of asymmetry, $(C_0=0)$} For sample-a in Table 1, POPC GUVs in water, using Eq. \ref{tension}, for $C_0=0$, we get 
\begin{equation}
\label{smallt}
\sigma = \frac{\kappa \Sigma}{R^2} \approx (3.85 \pm 0.08) \ {k_BT}/\mu \rm{m}^2
\end{equation}
\noindent
where we examined at least 25 measurements on different GUVs from repeated experiments and took an average. This gives an average estimate of membrane tension for POPC GUVs in water in the absence of AA.
\paragraph{In the presence membrane asymmetry, $(C_0 \neq 0)$} 
We could observe membrane flicker for all GUVs for  AA concentrations 0.14$\mu$M, 1.02$\mu$M, 1.19$\mu$M and 1.34$\mu$M. If we increase the concentration of AA above $\approx 1.5 \ \mu$M, then it appears that the membrane is favorable for bending for the formation of outbuds as shown in Figure \ref{varyc}. Above $18 \ \mu$M $(\approx$ 1 mg/ml) of AA concentration the shape of GUVs collapse/shrink completely suggesting a highly convoluted state. All vesicles would be subjected to the same concentration of AA within the experiment. Now, for an outward bend of the GUV membrane shown in Figures \ref{varyc} and \ref{budding}, the outer and inner leaflets of the bilayer undergo expansion and compression, respectively, relative to the mid surface, first emphasized by Evans.\cite{b58} \\
The presence of outbuds in Figure \ref{budding} directly demonstrates that AA has a preferable interaction with the membrane, giving only a positive spontaneous curvature of $C_0 \leq 0.05  \ \mu \rm{m}^{-1}$ at $18 \ \mu$M $(\approx$ 1 mg/ml) of AA concentration. 
\noindent
Vesicles shown in figure \ref{varyc} change their morphology due to increased AA concentration in the exterior of the vesicles. The outward budding is clearly visible and Figures \ref{varyc}a-d show the response of GUVs to the addition of AA with concentrations (a) 0 $\mu$M (b) 5 $\mu$M (c) 9.4 $\mu$M (d) 18 $\mu$M, respectively. The scale bars are $10 \ \mu m$. The GUVs shown in Figures \ref{varyc} b-d have several curved structures, which are smaller buds formed during the enzyme-GUV interaction. An example of time-dependent shape changes driven by the membrane asymmetry generated by AA is provided in Figures \ref{budding}a-c. In response to this asymmetry, the same GUV over time shows a budding transition due to the addition of AA at $18 \ \mu$M. We have shown a GUV in the absence of AA in Figure \ref{varyc}a. However, given the dynamic nature of the morphological change process, the focus on one plane was difficult for static imaging. In Figure \ref{budding}, the GUV shows changes in shape and remains in focus throughout the imaging process. The budding of vesicles in the living cell is a well-known phenomenon. \cite{b52} Previous studies have shown that small solutes, such as sugars, that are in contact with lipid membranes, can induce membrane curvature by forming layers of adsorption or depletion on both sides of the bilayer membrane.\cite{b59,b60,b57} 
\section*{Discussion}\label{sec8}
In this work we have studied the effect of soluble AA in pure water solution on the bending rigidity of POPC membranes. A relative complex picture emerge from our data. At low concentrations below $1.5 \ \mu$M of AA, $\kappa$ increases substantially, which is notable since the protein is not expected to be membrane associated. The surface groups of the protein does not suggest membrane binding, although the presence of some aromatic moieties can give some membrane interfacial affinity. At higher concentrations $\kappa$ is decreasing but remains higher than the value estimated for POPC in water in the investigated range up $18 \ \mu$M of AA. \\
It is well established that proteins can modify the mechanical properties of membranes, e.g. peripheral proteins can locally induce a spontaneous curvature, leading to a reduction in the effective bending rigidity $\kappa$ predicted theoretically  \cite{b64} and found experimentally for numerous systems (reviewed in \cite{b3}). Furthermore, it has been shown that integral proteins can increase the effective $\kappa$ of the membrane, for example, \cite{b65}. However, the AA soluble protein does not fall into these categories. \\
The presence of AA in solution significantly increases the stiffness of POPC lipid bilayer membrane bending, comparable to the effect of cholesterol, a well-known membrane stiffening compound, in the membrane (Figure 8). Numerous soluble compounds have been shown to decrease $\kappa$ \cite{b3} e.g., Hepes Buffer as shown in Figure 8.  Our data also show that the interaction between AA and the outer bilayer leaflet results in a positive spontanous curvature as shown in Figure 10. A possible explanation is that a  small affinity to the membrane, e.g. though surface tryptophans at the protein, is causing a net positive surface charge and thereby an increase in effective $\kappa$. \cite{b81,b82,b83} Such effects have been demontrated experimentally for lipidated charged peptides \cite{b84} where the surface charge is self-regulated.  Here, the electrostatic bending rigidity increase to a maximum  at low peptide concentrations in bulk solution while it is declining as the peptide concentration is increasing due to screening effects. A similar phenomena is a possiblity for POPC membranes and AA in pure water solvent, although the charge screening  is  less effective for the multivalent proteins and it counter ions and no added salt. We have demonstrated that the presence of soluble AA interact with the zwitteionic lipid membrane and change it's mechanical properties. \\
Soluble AA are of great interest to biology and medicine due to its many applications e.g. in fermentation processes, wound healing, brewing, baking, textile treatments, sewage treatment, detergents, etc. \cite{b71,b55,a1,a2,a3,a10}. The development of a microbial biofilm on the surface of the wound is always the main cause of wound healing failure. \cite{b80} EPS matrix forms the integrated structure of the biofilm and prevents penetration of antibiotics into the microbial cells. Thus, the application of components that can target / break down / dissolve biofilms would make them an attractive strategy for wound healing.\cite{b80,b78} 
\subsection*{Materials and Methods}
All commercially available reagents were used without any further purification. The solutions were prepared in Milli-Q water. 
\subsection*{Lipids and GUV preparation}
1-palmitoyl-2-oleoyl-sn-glycero-3-phosphocholine (POPC) (catalog number 850457P), and cholesterol (catalog number C8503) lipids were obtained from Sigma. Lipids are dissolved in chloroform such that the final concentration of the lipid is 1 mg/ml. The electroformation method was used to prepare GUVs in different hydration solutions at room temperature at 25 ° C.  Approximately $10 \ \mu l$ of the lipid solution was spread on each of two glass plates coated with indium tin oxide, which was left in the desiccator under vacuum for 45 minutes. The lipid film was electroswelled in the respective hydration solution at room temperature using a sinusoidal electric field generated using a function generator (Tektronix AFG3022C). We prepare POPC GUVs in Milli-Q and 10mM HEPES, POPC-cholesterol (10)\%  GUVs in 10mM sucrose. Amylase from malt was obtained from Himedia (catalog number GRM638). It has a molecular weight of 56 kDa and a polydispersity index of 0.276. Barley $\alpha$ -amylases have an overall ellipsoidal shape with approximate principal dimensions $6.8 \mathrm{nm}\times
5.3 \mathrm{nm}\times  3.6 \mathrm{nm}$. \cite{b73} corresponding to an excluding volume about $68\ \mathrm{nm^3}$.
\subsection*{Fluctuation Data Collection}
GUVs were observed using phase contrast microscopy (Zeiss Axio Observer 5) equipped with a 20x objective (PL APO 20x/0.8 Ph2 M27). The objective has a working distance of 0.55 mm and a field of view of 25 mm. All images and movies were recorded with a 20X objective using a Zeiss Axiocam 506 mono camera. About 25,000 frames have been recorded for a single GUV at an exposure time of 29 fps. 
\subsection*{GUV experiments with the enzyme}
A stock solution of 10 mg/ml $(178.6 \ \mu$M) of AA (56 kDa) was prepared. To verify the effect of AA on the stiffness of biomembranes, POPC GUVs with varying concentrations of AA were incubated, giving a final concentration as indicated in Table 1.
\subsection*{Acknowledgement}
This research was carried out at the Indian Institute of Science Education and Research Mohali and is supported by the Ramalingaswami grant of DBT (BT / RLF / Re-entry grant / 06/2020). The authors declare that they have no competing interests.

\subsection*{Author contributions}
T.B. designed the project and supervised the experiments, and J.H.I. developed the theory and the flickering code. J.H.I. guided S.M. in the programming. H.K. did the AA experiments. S.G. and H.M. did the experiments without the AA. M.K. and R.K. helped with the microscopy experiments and prepared figures and additional information. P.S. and I.P.K. contributed AA. T.B. and J.H.I. wrote the manuscript. All authors discussed the results and commented on the manuscript.

% Bibliography
\bibliography{sn-article}


\begin{thebibliography}{51}
\ifx \bisbn   \undefined \def \bisbn  #1{ISBN #1}\fi
\ifx \binits  \undefined \def \binits#1{#1}\fi
\ifx \bauthor  \undefined \def \bauthor#1{#1}\fi
\ifx \batitle  \undefined \def \batitle#1{#1}\fi
\ifx \bjtitle  \undefined \def \bjtitle#1{#1}\fi
\ifx \bvolume  \undefined \def \bvolume#1{\textbf{#1}}\fi
\ifx \byear  \undefined \def \byear#1{#1}\fi
\ifx \bissue  \undefined \def \bissue#1{#1}\fi
\ifx \bfpage  \undefined \def \bfpage#1{#1}\fi
\ifx \blpage  \undefined \def \blpage #1{#1}\fi
\ifx \burl  \undefined \def \burl#1{\textsf{#1}}\fi
\ifx \doiurl  \undefined \def \doiurl#1{\url{https://doi.org/#1}}\fi
\ifx \betal  \undefined \def \betal{\textit{et al.}}\fi
\ifx \binstitute  \undefined \def \binstitute#1{#1}\fi
\ifx \binstitutionaled  \undefined \def \binstitutionaled#1{#1}\fi
\ifx \bctitle  \undefined \def \bctitle#1{#1}\fi
\ifx \beditor  \undefined \def \beditor#1{#1}\fi
\ifx \bpublisher  \undefined \def \bpublisher#1{#1}\fi
\ifx \bbtitle  \undefined \def \bbtitle#1{#1}\fi
\ifx \bedition  \undefined \def \bedition#1{#1}\fi
\ifx \bseriesno  \undefined \def \bseriesno#1{#1}\fi
\ifx \blocation  \undefined \def \blocation#1{#1}\fi
\ifx \bsertitle  \undefined \def \bsertitle#1{#1}\fi
\ifx \bsnm \undefined \def \bsnm#1{#1}\fi
\ifx \bsuffix \undefined \def \bsuffix#1{#1}\fi
\ifx \bparticle \undefined \def \bparticle#1{#1}\fi
\ifx \barticle \undefined \def \barticle#1{#1}\fi
\bibcommenthead
\ifx \bconfdate \undefined \def \bconfdate #1{#1}\fi
\ifx \botherref \undefined \def \botherref #1{#1}\fi
\ifx \url \undefined \def \url#1{\textsf{#1}}\fi
\ifx \bchapter \undefined \def \bchapter#1{#1}\fi
\ifx \bbook \undefined \def \bbook#1{#1}\fi
\ifx \bcomment \undefined \def \bcomment#1{#1}\fi
\ifx \oauthor \undefined \def \oauthor#1{#1}\fi
\ifx \citeauthoryear \undefined \def \citeauthoryear#1{#1}\fi
\ifx \endbibitem  \undefined \def \endbibitem {}\fi
\ifx \bconflocation  \undefined \def \bconflocation#1{#1}\fi
\ifx \arxivurl  \undefined \def \arxivurl#1{\textsf{#1}}\fi
\csname PreBibitemsHook\endcsname

%%% 1
\bibitem{b77}
\begin{barticle}
\bauthor{\bsnm{Schenkels}, \binits{L.C.}},
\bauthor{\bsnm{Veerman}, \binits{E.C.}},
\bauthor{\bsnm{Amerongen}, \binits{A.V.N.}}:
\batitle{Biochemical composition of human saliva in relation to other mucosal
  fluids.}
\bjtitle{Crit. Rev. Oral Biol.}
\bvolume{6},
\bfpage{161}--\blpage{175}
(\byear{1995})
\end{barticle}
\endbibitem

%%% 2
\bibitem{b76}
\begin{barticle}
\bauthor{\bsnm{Dawes}, \binits{C.}},
\bauthor{\bsnm{Pedersen}, \binits{A.}},
\bauthor{\bsnm{Villa}, \binits{A.}},
\bauthor{\bsnm{Ekström}, \binits{J.}},
\bauthor{\bsnm{Proctor}, \binits{G.B.}},
\bauthor{\bsnm{Vissink}, \binits{A.}},
\bauthor{\bsnm{Aframian}, \binits{D.}},
\bauthor{\bsnm{McGowan}, \binits{R.}},
\bauthor{\bsnm{Aliko}, \binits{A.}},
\bauthor{\bsnm{Narayana}, \binits{N.}},
\bauthor{\bsnm{Sia}, \binits{Y.W.}},
\bauthor{\bsnm{Joshi}, \binits{R.K.}},
\bauthor{\bsnm{Jensen}, \binits{S.B.}},
\bauthor{\bsnm{Kerr}, \binits{A.R.}},
\bauthor{\bsnm{Wolff}, \binits{A.}}:
\batitle{The functions of human saliva: A review sponsored by the world
  workshop on oral medicine vi.}
\bjtitle{Arch. Oral Biol.}
\bvolume{60},
\bfpage{863}--\blpage{874}
(\byear{2015})
\end{barticle}
\endbibitem

%%% 3
\bibitem{b78}
\begin{barticle}
\bauthor{\bsnm{Vaikundamoorthy}, \binits{R.}},
\bauthor{\bsnm{Rajendran}, \binits{R.}},
\bauthor{\bsnm{Selvaraju}, \binits{A.}},
\bauthor{\bsnm{Moorthy}, \binits{K.}},
\bauthor{\bsnm{Perumal}, \binits{S.}}:
\batitle{Development of thermostable amylase enzyme from bacillus cereus for
  potential antibiofilm activity}.
\bjtitle{Bioorg. Chem.}
\bvolume{77},
\bfpage{494}--\blpage{506}
(\byear{2018})
\end{barticle}
\endbibitem

%%% 4
\bibitem{b79}
\begin{barticle}
\bauthor{\bsnm{Pallavicini}, \binits{P.}},
\bauthor{\bsnm{Arciola}, \binits{C.R.}},
\bauthor{\bsnm{Bertoglio}, \binits{F.}},
\bauthor{\bsnm{Curtosi}, \binits{S.}},
\bauthor{\bsnm{Dacarro}, \binits{G.}},
\bauthor{\bsnm{D'Agostino}, \binits{A.}},
\bauthor{\bsnm{Ferrari}, \binits{F.}},
\bauthor{\bsnm{Merli}, \binits{D.}},
\bauthor{\bsnm{Milanese}, \binits{C.}},
\bauthor{\bsnm{Rossi}, \binits{S.}},
\bauthor{\bsnm{Taglietti}, \binits{A.}},
\bauthor{\bsnm{Tenci}, \binits{M.}},
\bauthor{\bsnm{Visai}, \binits{L.}}:
\batitle{Silver nanoparticles synthesized and coated with pectin: An ideal
  compromise for anti-bacterial and anti-biofilm action combined with
  wound-healing properties}.
\bjtitle{J. Colloid Interface Sci.}
\bvolume{498},
\bfpage{271}--\blpage{281}
(\byear{2017})
\end{barticle}
\endbibitem

%%% 5
\bibitem{b70}
\begin{barticle}
\bauthor{\bsnm{Mäntsälä}, \binits{P.}},
\bauthor{\bsnm{Zalkin}, \binits{H.}}:
\batitle{Membrane-bound and soluble extracellular alpha-amylase from bacillus
  subtilis}.
\bjtitle{J. Biol. Chem}
\bvolume{254},
\bfpage{8540}--\blpage{8547}
(\byear{1979})
\end{barticle}
\endbibitem

%%% 6
\bibitem{b72}
\begin{barticle}
\bauthor{\bsnm{Olsen}, \binits{S.N.}},
\bauthor{\bsnm{Andersen}, \binits{K.B.}},
\bauthor{\bsnm{Randolph}, \binits{T.W.}},
\bauthor{\bsnm{Carpenter}, \binits{J.F.}},
\bauthor{\bsnm{Westh}, \binits{P.}}:
\batitle{Role of electrostatic repulsion on colloidal stability of bacillus
  halmapalus alpha-amylase}.
\bjtitle{Biochimica et Biophysica Acta (BBA)-Proteins and Proteomics}
\bvolume{1794},
\bfpage{1058}--\blpage{1065}
(\byear{2009})
\end{barticle}
\endbibitem

%%% 7
\bibitem{b74}
\begin{barticle}
\bauthor{\bsnm{Warren}, \binits{F.J.}},
\bauthor{\bsnm{G.}, \binits{P.}},
\bauthor{\bsnm{Royall}, \binits{P.G.}},
\bauthor{\bsnm{S.}, \binits{G.}},
\bauthor{\bsnm{Butterworth}, \binits{P.J.}},
\bauthor{\bsnm{Ellis}, \binits{P.R.}}:
\batitle{Binding interactions of alpha-amylase with starch granules: The
  influence of supramolecular structure and surface area}.
\bjtitle{Carbohydrate polymers}
\bvolume{86},
\bfpage{1038}--\blpage{1047}
(\byear{2011})
\end{barticle}
\endbibitem

%%% 8
\bibitem{b75}
\begin{botherref}
\oauthor{\bsnm{Baroroh}, \binits{U.}},
\oauthor{\bsnm{Yusuf}, \binits{M.}},
\oauthor{\bsnm{Rachman}, \binits{S.D.}},
\oauthor{\bsnm{Ishmayana}, \binits{S.}},
\oauthor{\bsnm{Syamsunarno}, \binits{M.R.A.A.}},
\oauthor{\bsnm{Levita}, \binits{J.}},
\oauthor{\bsnm{Subroto}, \binits{T.}}:
The importance of surface-binding site towards starch-adsorptivity level in
  alpha-amylase: a review on structural point of view.
Enzyme research
\textbf{2017}
(2017)
\end{botherref}
\endbibitem

%%% 9
\bibitem{b10}
\begin{barticle}
\bauthor{\bsnm{Helfrich}, \binits{W.}}:
\batitle{Elastic properties of lipid bilayers: theory and possible
  experiments}.
\bjtitle{Zeitschrift f{\"u}r Naturforschung c}
\bvolume{28}(\bissue{11-12}),
\bfpage{693}--\blpage{703}
(\byear{1973})
\end{barticle}
\endbibitem

%%% 10
\bibitem{b0}
\begin{barticle}
\bauthor{\bsnm{Brochard}, \binits{F.}},
\bauthor{\bsnm{Lennon}, \binits{J.F.}}:
\batitle{Frequency spectrum of the flicker phenomenon in erythrocytes}.
\bjtitle{Journal de Physique}
\bvolume{36}(\bissue{11}),
\bfpage{1035}--\blpage{1047}
(\byear{1975})
\end{barticle}
\endbibitem

%%% 11
\bibitem{b31}
\begin{barticle}
\bauthor{\bsnm{Marsh}, \binits{D.}}:
\batitle{Elastic curvature constants of lipid monolayers and bilayers}.
\bjtitle{Chemistry and Physics of Lipids}
\bvolume{144}(\bissue{2}),
\bfpage{146}--\blpage{159}
(\byear{2006})
\end{barticle}
\endbibitem

%%% 12
\bibitem{b18}
\begin{barticle}
\bauthor{\bsnm{Faucon}, \binits{J.F.}},
\bauthor{\bsnm{Mitov}, \binits{M.D.}},
\bauthor{\bsnm{M{\'e}l{\'e}ard}, \binits{P.}},
\bauthor{\bsnm{Bivas}, \binits{I.}},
\bauthor{\bsnm{Bothorel}, \binits{P.}}:
\batitle{Bending elasticity and thermal fluctuations of lipid membranes.
  theoretical and experimental requirements}.
\bjtitle{Journal de physique}
\bvolume{50}(\bissue{17}),
\bfpage{2389}--\blpage{2414}
(\byear{1989})
\end{barticle}
\endbibitem

%%% 13
\bibitem{a7}
\begin{barticle}
\bauthor{\bsnm{Matsuura}, \binits{Y.}},
\bauthor{\bsnm{Kusunoki}, \binits{M.}},
\bauthor{\bsnm{Harada}, \binits{W.}},
\bauthor{\bsnm{Kakudo}, \binits{M.}}:
\batitle{Structure and possible catalytic residues of taka-amylase a}.
\bjtitle{J. Biochem.}
\bvolume{95}(\bissue{3}),
\bfpage{697}--\blpage{702}
(\byear{1984})
\end{barticle}
\endbibitem

%%% 14
\bibitem{b43}
\begin{barticle}
\bauthor{\bsnm{D{\"o}bereiner}, \binits{H.G.}},
\bauthor{\bsnm{Gompper}, \binits{G.}},
\bauthor{\bsnm{Haluska}, \binits{C.K.}},
\bauthor{\bsnm{Kroll}, \binits{D.M.}},
\bauthor{\bsnm{Petrov}, \binits{P.G.}},
\bauthor{\bsnm{Riske}, \binits{K.A.}}:
\batitle{Advanced flicker spectroscopy of fluid membranes}.
\bjtitle{Phys. Rev. Lett}
\bvolume{91}(\bissue{4}),
\bfpage{048301}
(\byear{2003})
\end{barticle}
\endbibitem

%%% 15
\bibitem{b44}
\begin{barticle}
\bauthor{\bsnm{Engelhardt}, \binits{H.}},
\bauthor{\bsnm{Duwe}, \binits{H.P.}},
\bauthor{\bsnm{Sackmann}, \binits{E.}}:
\batitle{Bilayer bending elasticity measured by fourier analysis of thermally
  excited surface undulations of flaccid vesicles}.
\bjtitle{Journal de Physique Letters}
\bvolume{46}(\bissue{8}),
\bfpage{395}--\blpage{400}
(\byear{1985})
\end{barticle}
\endbibitem

%%% 16
\bibitem{b47}
\begin{barticle}
\bauthor{\bsnm{Cai}, \binits{W.}},
\bauthor{\bsnm{Lubensky}, \binits{T.C.}},
\bauthor{\bsnm{Nelson}, \binits{P.C.}},
\bauthor{\bsnm{Powers}, \binits{T.}}:
\batitle{Measure factors, tension, and correlations of fluid membranes}.
\bjtitle{Journal De Physique II}
\bvolume{4}(\bissue{6}),
\bfpage{931}--\blpage{949}
(\byear{1994})
\end{barticle}
\endbibitem

%%% 17
\bibitem{b33}
\begin{barticle}
\bauthor{\bsnm{Pezeshkian}, \binits{W.}},
\bauthor{\bsnm{Ipsen}, \binits{J.}}:
\batitle{Fluctuations and conformational stability of a membrane patch with
  curvature inducing inclusions}.
\bjtitle{Soft Matter}
\bvolume{15}(\bissue{48}),
\bfpage{9974}--\blpage{81}
(\byear{2019}).
\doiurl{10.1039/C9SM01762C}
\end{barticle}
\endbibitem

%%% 18
\bibitem{b23}
\begin{barticle}
\bauthor{\bsnm{Genova}, \binits{J.}},
\bauthor{\bsnm{Zheliaskova}, \binits{A.}},
\bauthor{\bsnm{Mitov}, \binits{M.D.}}:
\batitle{The influence of sucrose on the elasticity of sopc lipid membrane
  studied by the analysis of thermally induced shape fluctuations}.
\bjtitle{Colloids and Surfaces A: Physicochemical and Engineering Aspects}
\bvolume{282},
\bfpage{420}--\blpage{422}
(\byear{2006})
\end{barticle}
\endbibitem

%%% 19
\bibitem{b39}
\begin{barticle}
\bauthor{\bsnm{Bouvrais}, \binits{H.}},
\bauthor{\bsnm{Duelund}, \binits{L.}},
\bauthor{\bsnm{Ipsen}, \binits{J.H.}}:
\batitle{Buffers affect the bending rigidity of model lipid membranes}.
\bjtitle{Langmuir}
\bvolume{30}(\bissue{1}),
\bfpage{13}--\blpage{16}
(\byear{2014})
\end{barticle}
\endbibitem

%%% 20
\bibitem{b35}
\begin{barticle}
\bauthor{\bsnm{Henriksen}, \binits{J.}},
\bauthor{\bsnm{Rowat}, \binits{A.C.}},
\bauthor{\bsnm{Brief}, \binits{E.}},
\bauthor{\bsnm{Hsueh}, \binits{Y.W.}},
\bauthor{\bsnm{Thewalt}, \binits{J.L.}},
\bauthor{\bsnm{Zuckermann}, \binits{M.J.}},
\bauthor{\bsnm{Ipsen}, \binits{J.H.}}:
\batitle{Universal behavior of membranes with sterols}.
\bjtitle{Biophys. J.}
\bvolume{90}(\bissue{5}),
\bfpage{1639}--\blpage{1649}
(\byear{2006})
\end{barticle}
\endbibitem

%%% 21
\bibitem{b40}
\begin{barticle}
\bauthor{\bsnm{Servuss}, \binits{R.M.}},
\bauthor{\bsnm{Harbich}, \binits{V.}},
\bauthor{\bsnm{Helfrich}, \binits{W.}}:
\batitle{Measurement of the curvature-elastic modulus of egg lecithin
  bilayers}.
\bjtitle{Biochim. Biophys. Acta-Biomembranes}
\bvolume{436}(\bissue{4}),
\bfpage{900}--\blpage{903}
(\byear{1976})
\end{barticle}
\endbibitem

%%% 22
\bibitem{b45}
\begin{barticle}
\bauthor{\bsnm{Duwe}, \binits{H.P.}},
\bauthor{\bsnm{Kaes}, \binits{J.}},
\bauthor{\bsnm{Sackmann}, \binits{E.}}:
\batitle{Bending elastic moduli of lipid bilayers : modulation by solutes}.
\bjtitle{Journal de Physique}
\bvolume{51}(\bissue{10}),
\bfpage{945}--\blpage{961}
(\byear{1990}).
\doiurl{10.1051/jphys:019900051010094500}
\end{barticle}
\endbibitem

%%% 23
\bibitem{b42}
\begin{barticle}
\bauthor{\bsnm{Beblik}, \binits{G.}},
\bauthor{\bsnm{Servuss}, \binits{R.M.}},
\bauthor{\bsnm{Helfrich}, \binits{W.}}:
\batitle{Bilayer bending rigidity of some synthetic lecithins}.
\bjtitle{Journal de Physique}
\bvolume{46}(\bissue{10}),
\bfpage{1773}--\blpage{1778}
(\byear{1985})
\end{barticle}
\endbibitem

%%% 24
\bibitem{b41}
\begin{barticle}
\bauthor{\bsnm{Schneider}, \binits{M.B.}},
\bauthor{\bsnm{Jenkins}, \binits{J.T.}},
\bauthor{\bsnm{Webb}, \binits{W.W.}}:
\batitle{Thermal fluctuations of large quasi-spherical bimolecular phospholipid
  vesicles}.
\bjtitle{Journal de Physique}
\bvolume{45}(\bissue{9}),
\bfpage{1457}--\blpage{1472}
(\byear{1984})
\end{barticle}
\endbibitem

%%% 25
\bibitem{b37}
\begin{barticle}
\bauthor{\bsnm{Bouvrais}, \binits{H.}},
\bauthor{\bsnm{Pott}, \binits{T.}},
\bauthor{\bsnm{Bagatolli}, \binits{L.A.}},
\bauthor{\bsnm{Ipsen}, \binits{J.H.}},
\bauthor{\bsnm{M{\'e}l{\'e}ard}, \binits{P.}}:
\batitle{Impact of membrane-anchored fluorescent probes on the mechanical
  properties of lipid bilayers}.
\bjtitle{Biochim. Biophys. Acta-Biomembranes}
\bvolume{1798}(\bissue{7}),
\bfpage{1333}--\blpage{1337}
(\byear{2010})
\end{barticle}
\endbibitem

%%% 26
\bibitem{b36}
\begin{barticle}
\bauthor{\bsnm{Bouvrais}, \binits{H.}},
\bauthor{\bsnm{M{\'e}l{\'e}ard}, \binits{P.}},
\bauthor{\bsnm{Pott}, \binits{T.}},
\bauthor{\bsnm{Jensen}, \binits{K.J.}},
\bauthor{\bsnm{Brask}, \binits{J.}},
\bauthor{\bsnm{Ipsen}, \binits{J.H.}}:
\batitle{Softening of popc membranes by magainin}.
\bjtitle{Biophysical chemistry}
\bvolume{137}(\bissue{1}),
\bfpage{7}--\blpage{12}
(\byear{2008})
\end{barticle}
\endbibitem

%%% 27
\bibitem{b34}
\begin{barticle}
\bauthor{\bsnm{Niggemann}, \binits{G.}},
\bauthor{\bsnm{Kummrow}, \binits{M.}},
\bauthor{\bsnm{Helfrich}, \binits{W.}}:
\batitle{The bending rigidity of phosphatidylcholine bilayers: dependences on
  experimental method, sample cell sealing and temperature}.
\bjtitle{Journal de Physique II}
\bvolume{5}(\bissue{3}),
\bfpage{413}--\blpage{425}
(\byear{1995})
\end{barticle}
\endbibitem

%%% 28
\bibitem{b9}
\begin{barticle}
\bauthor{\bsnm{Henriksen}, \binits{J.}},
\bauthor{\bsnm{Rowat}, \binits{A.C.}},
\bauthor{\bsnm{Ipsen}, \binits{J.H.}}:
\batitle{Vesicle fluctuation analysis of the effects of sterols on membrane
  bending rigidity}.
\bjtitle{Eur. Biophys. J.}
\bvolume{33},
\bfpage{732}--\blpage{741}
(\byear{2004})
\end{barticle}
\endbibitem

%%% 29
\bibitem{b3}
\begin{bchapter}
\bauthor{\bsnm{Ipsen}, \binits{J.H.}},
\bauthor{\bsnm{Hansen}, \binits{A.G.}},
\bauthor{\bsnm{Bhatia}, \binits{T.}}:
\bctitle{Vesicle fluctuation analysis}.
In: \bbtitle{The Giant Vesicle Book},
pp. \bfpage{333}--\blpage{345}.
\bpublisher{CRC Press},
\blocation{Boca Raton}
(\byear{2019})
\end{bchapter}
\endbibitem

%%% 30
\bibitem{b57}
\begin{barticle}
\bauthor{\bsnm{Bhatia}, \binits{T.}},
\bauthor{\bsnm{Christ}, \binits{S.}},
\bauthor{\bsnm{Steinkühler}, \binits{J.}},
\bauthor{\bsnm{Dimova}, \binits{R.}},
\bauthor{\bsnm{Lipowsky}, \binits{R.}}:
\batitle{Simple sugars shape giant vesicles into multispheres with many
  membrane necks}.
\bjtitle{Soft Matter}
\bvolume{16},
\bfpage{1246}--\blpage{1258}
(\byear{2020}).
\doiurl{10.1039/C9SM01890E}
\end{barticle}
\endbibitem

%%% 31
\bibitem{b59}
\begin{barticle}
\bauthor{\bsnm{Nowbagh}, \binits{A.}},
\bauthor{\bsnm{Deshwal}, \binits{A.}},
\bauthor{\bsnm{Kadu}, \binits{M.}},
\bauthor{\bsnm{Chaudhuri}, \binits{A.}},
\bauthor{\bsnm{Maiti}, \binits{S.}},
\bauthor{\bsnm{Lipowsky}, \binits{R.}},
\bauthor{\bsnm{Bhatia}, \binits{T.}}:
\batitle{Generation of bilayer asymmetry and membrane curvature by the
  sugar‐cleaving enzyme invertase}.
\bjtitle{ChemSystemsChem}
\bvolume{5},
\bfpage{202200027}
(\byear{2022})
\end{barticle}
\endbibitem

%%% 32
\bibitem{b60}
\begin{barticle}
\bauthor{\bsnm{Bhatia}, \binits{T.}}:
\batitle{Micromechanics of biomembranes}.
\bjtitle{J. Membr. Biol.}
\bvolume{255},
\bfpage{637}--\blpage{649}
(\byear{2022})
\end{barticle}
\endbibitem

%%% 33
\bibitem{b69}
\begin{barticle}
\bauthor{\bsnm{Lipowsky}, \binits{R.}}:
\batitle{Remodeling of membrane shape and topology by curvature elasticity and
  membrane tension}.
\bjtitle{Advanced Biology}
\bvolume{6},
\bfpage{2101020}
(\byear{2022})
\end{barticle}
\endbibitem

%%% 34
\bibitem{b50}
\begin{barticle}
\bauthor{\bsnm{Rothman}, \binits{J.E.}},
\bauthor{\bsnm{Lenard}, \binits{J.}}:
\batitle{Membrane asymmetry : The nature of membrane asymmetry provides clues
  to the puzzle of how membranes are assembled}.
\bjtitle{Science}
\bvolume{195}(\bissue{4280}),
\bfpage{743}--\blpage{753}
(\byear{1977})
\end{barticle}
\endbibitem

%%% 35
\bibitem{b51}
\begin{barticle}
\bauthor{\bsnm{Lodish}, \binits{H.F.}},
\bauthor{\bsnm{Rothman}, \binits{J.E.}}:
\batitle{The assembly of cell membranes}.
\bjtitle{Scientific American}
\bvolume{240}(\bissue{1}),
\bfpage{48}--\blpage{63}
(\byear{1979})
\end{barticle}
\endbibitem

%%% 36
\bibitem{b58}
\begin{barticle}
\bauthor{\bsnm{Evans}, \binits{E.}}:
\batitle{Bending resistance and chemically induced moments in membrane
  bilayers}.
\bjtitle{Biophys. J.}
\bvolume{14},
\bfpage{923}--\blpage{31}
(\byear{1974})
\end{barticle}
\endbibitem

%%% 37
\bibitem{b52}
\begin{barticle}
\bauthor{\bsnm{Rothman}, \binits{J.E.}},
\bauthor{\bsnm{Orci}, \binits{L.}}:
\batitle{Budding vesicles in living cells}.
\bjtitle{Scientific American}
\bvolume{274}(\bissue{3}),
\bfpage{70}--\blpage{75}
(\byear{1996})
\end{barticle}
\endbibitem

%%% 38
\bibitem{b64}
\begin{barticle}
\bauthor{\bsnm{Netz}, \binits{R.R.}},
\bauthor{\bsnm{Pincus}, \binits{P.}}:
\batitle{Inhomogeneous fluid membranes: Segregation, ordering, and effective
  rigidity}.
\bjtitle{Phys. Rev. E}
\bvolume{52},
\bfpage{4114}
(\byear{1995})
\end{barticle}
\endbibitem

%%% 39
\bibitem{b65}
\begin{barticle}
\bauthor{\bsnm{Bouvrais}, \binits{H.}},
\bauthor{\bsnm{Cornelius}, \binits{F.}},
\bauthor{\bsnm{Ipsen}, \binits{J.H.}},
\bauthor{\bsnm{Mouritsen}, \binits{O.G.}}:
\batitle{Intrinsic reaction-cycle time scale of na+,k+-atpase manifests itself
  in the lipid-protein interactions of nonequilibrium membranes}.
\bjtitle{Proc. Natl. Acad. Sci. U. S. A.}
\bvolume{109},
\bfpage{18442}--\blpage{6}
(\byear{2012})
\end{barticle}
\endbibitem

%%% 40
\bibitem{b81}
\begin{barticle}
\bauthor{\bsnm{Winterhalter}, \binits{M.}},
\bauthor{\bsnm{H.}, \binits{W.}}:
\batitle{Effect of surface charge on the curvature elasticity of membranes.}
\bjtitle{J. Phys. Chem.}
\bvolume{92},
\bfpage{6865}--\blpage{6867}
(\byear{1988})
\end{barticle}
\endbibitem

%%% 41
\bibitem{b82}
\begin{barticle}
\bauthor{\bsnm{Fogden}, \binits{A.}},
\bauthor{\bsnm{Mitchell}, \binits{D.J.}},
\bauthor{},
\bauthor{\bsnm{Ninham}, \binits{B.W.}}:
\batitle{Undulations of charged membranes.}
\bjtitle{Langmuir}
\bvolume{6},
\bfpage{159}--\blpage{162}
(\byear{1990})
\end{barticle}
\endbibitem

%%% 42
\bibitem{b83}
\begin{barticle}
\bauthor{\bsnm{Lekkerkerker}, \binits{H.N.W.}}:
\batitle{Contribution of the electric double layer to the curvature elasticity
  of charged amphiphilic monolayers.}
\bjtitle{Physica A: Statistical Mechanics and its Applications}
\bvolume{159},
\bfpage{319}--\blpage{328}
(\byear{1989})
\end{barticle}
\endbibitem

%%% 43
\bibitem{b84}
\begin{barticle}
\bauthor{\bsnm{Rowat}, \binits{A.C.}},
\bauthor{\bsnm{Hansen}, \binits{P.L.}},
\bauthor{\bsnm{Ipsen}, \binits{J.H.}}:
\batitle{Experimental evidence of the electrostatic contribution to membrane
  bending rigidity.}
\bjtitle{EPL}
\bvolume{67},
\bfpage{144}
(\byear{2004})
\end{barticle}
\endbibitem

%%% 44
\bibitem{b71}
\begin{barticle}
\bauthor{\bsnm{Mauno}, \binits{V.}},
\bauthor{\bsnm{Mäntsälä}, \binits{P.}}:
\batitle{Microbial amylolytic enzyme}.
\bjtitle{Critical reviews in biochemistry and molecular biology}
\bvolume{24},
\bfpage{329}--\blpage{418}
(\byear{1989})
\end{barticle}
\endbibitem

%%% 45
\bibitem{b55}
\begin{barticle}
\bauthor{\bsnm{Kirk}, \binits{O.}},
\bauthor{\bsnm{Borchert}, \binits{T.V.}},
\bauthor{\bsnm{Fuglsang}, \binits{C.C.}}:
\batitle{Industrial enzyme applications}.
\bjtitle{Current Opinion in Biotechnology}
\bvolume{13}(\bissue{4}),
\bfpage{345}--\blpage{351}
(\byear{2002}).
\doiurl{10.1016/S0958-1669(02)00328-2}
\end{barticle}
\endbibitem

%%% 46
\bibitem{a1}
\begin{barticle}
\bauthor{\bsnm{Azzopardi}, \binits{E.}},
\bauthor{\bsnm{C.}, \binits{L.}},
\bauthor{\bsnm{Teixeira}, \binits{S.R.}},
\bauthor{\bsnm{Conlan}, \binits{R.S.}},
\bauthor{\bsnm{Whitaker}, \binits{I.S.}}:
\batitle{Clinical applications of amylase: Novel perspectives}.
\bjtitle{Surgery}
\bvolume{160}(\bissue{1}),
\bfpage{26}--\blpage{37}
(\byear{2016}).
\doiurl{10.1016/j.surg.2016.01.005}
\end{barticle}
\endbibitem

%%% 47
\bibitem{a2}
\begin{barticle}
\bauthor{\bsnm{Boehlke}, \binits{C.}},
\bauthor{\bsnm{Zierau}, \binits{O.}},
\bauthor{\bsnm{Hannig}, \binits{C.}}:
\batitle{Salivary amylase - the enzyme of unspecialized euryphagous animals}.
\bjtitle{Arch Oral Biol.}
\bvolume{60}(\bissue{8}),
\bfpage{1162}--\blpage{76}
(\byear{2015}).
\doiurl{10.1016/j.archoralbio.2015.05.008}
\end{barticle}
\endbibitem

%%% 48
\bibitem{a3}
\begin{barticle}
\bauthor{\bsnm{Mondal}, \binits{S.}},
\bauthor{\bsnm{Mondal}, \binits{K.}},
\bauthor{\bsnm{Halder}, \binits{S.K.}},
\bauthor{\bsnm{Thakur}, \binits{N.}},
\bauthor{\bsnm{Mondal}, \binits{K.C.}}:
\batitle{Microbial amylase: Old but still at the forefront of all major
  industrial enzymes}.
\bjtitle{Biocatalysis and Agricultural Biotechnology}
\bvolume{45},
\bfpage{102509}
(\byear{2022}).
\doiurl{10.1016/j.archoralbio.2015.05.008}
\end{barticle}
\endbibitem

%%% 49
\bibitem{a10}
\begin{barticle}
\bauthor{\bsnm{Zhang}, \binits{Q.}},
\bauthor{\bsnm{Han}, \binits{Y.}},
\bauthor{\bsnm{Xiao}, \binits{H.}}:
\batitle{Microbial $\alpha$-amylase: A biomolecular overview}.
\bjtitle{Process Biochemistry}
\bvolume{53},
\bfpage{88}--\blpage{101}
(\byear{2017})
\end{barticle}
\endbibitem

%%% 50
\bibitem{b80}
\begin{botherref}
\oauthor{\bsnm{Phillips}, \binits{P.L.}},
\oauthor{\bsnm{Wolcott}, \binits{R.D.}},
\oauthor{\bsnm{Cowan}, \binits{L.J.}},
\oauthor{\bsnm{Schultz}, \binits{G.S.}}:
Biofilms in wounds and wound dressing.
In: Wound Healing Biomaterials
vol. 2
\end{botherref}
\endbibitem

%%% 51
\bibitem{b73}
\begin{barticle}
\bauthor{\bsnm{Robert}, \binits{X.}},
\bauthor{\bsnm{Haser}, \binits{R.}},
\bauthor{\bsnm{Gottschalk}, \binits{T.E.}},
\bauthor{\bsnm{Ratajczak}, \binits{F.}},
\bauthor{\bsnm{Driguez}, \binits{B.} \bsuffix{H.~Svensson}},
\bauthor{\bsnm{Aghajari}, \binits{N.}}:
\batitle{The structure of barley $\alpha$-amylase isozyme 1 reveals a novel
  role of domain c in substrate recognition and binding: a pair of sugar
  tongs.}
\bjtitle{Structure}
\bvolume{11},
\bfpage{973}--\blpage{984}
(\byear{2003})
\end{barticle}
\endbibitem

\end{thebibliography}

\end{document}